# Measurement of the Hyperfine Structure and Isotope Shifts of the $3s^2 3p^2$ $^3P_2 \rightarrow 3s3p^3$ $^3D^o_3$ Transition in Silicon

S. A. Lee[*] and W. M. Fairbank, Jr. [†]
Department of Physics
Colorado State University
Ft. Collins, CO  80523

**ABSTRACT**

The hyperfine structure and isotope shifts of the $3s^2 3p^2$ $^3P_2 \rightarrow 3s3p^3$ $^3D^o_3$ transition in silicon have been measured. The transition at 221.7 nm was studied by laser induced fluorescence in an atomic Si beam. For $^{29}Si$, the hyperfine A constant for the $3s^2 3p^2$ $^3P_2$ level was determined to be -160.1±1.3 MHz (1σ error), and the A constant for the $3s3p^3$ $^3D^o_3$ level is –532.9±0.6 MHz. This is the first time that these constants were measured. The isotope shifts (relative to the abundant isotope $^{28}Si$) of the transition were determined to be 1753.3±1.1 MHz for $^{29}Si$ and 3359.9±0.6 MHz for $^{30}Si$. This is an improvement by about two orders of magnitude over a previous measurement. From these results we are able to predict the hyperfine structure and isotope shift of the radioactive $^{31}Si$ atom, which is of interest in building a scalable quantum computer.

---

[*] Siuau.Lee@colostate.edu
[†] William.Fairbank@colostate.edu

I. INTRODUCTION

In 1998 Bruce Kane proposed a silicon-based quantum computer scheme that has the potential to be scalable to kilo-qubits or even mega-qubits [1]. It utilizes an array of single $^{31}$P dopant atoms embedded in a $^{28}$Si lattice. Each $^{31}$P atom has nuclear spin 1/2 and acts as a qubit. The formidable challenge for this concept is the placement of single $^{31}$P atoms, ~10 nm below the surface and separated by ~20 nm, with about 1 nm precision. Even the best commercial ion guns do not have sufficiently small phase space to accomplish this task at low enough deposition energy, ~100eV, to avoid >1 nm uncertainty in position due to ion straggling on sample entry. Deposition of one and only one ion in a site is also problematic with currently available sources. A laser cooled and trapped single atom in a magneto optic trap, followed by resonant photo-ionization, provides a deterministic ion source with a small enough phase space to achieve the required precision in deposition [2,3]. Current laser technology does not allow for the direct cooling and trapping of P atoms, as the transition at 177 nm is in the vacuum ultraviolet region. However, the $3s^23p^2\ ^3P_2 \rightarrow 3s3p^3\ ^3D^o_3$ transition in silicon at 221.7 nm is a cycling transition that may be used for the laser cooling and trapping of silicon atoms. In particular, the radioactive isotope $^{31}$Si would beta decay after deposition into $^{31}$P. Such a deterministic single ion source of $^{31}$Si could provide the desired $^{31}$P qubits for a scalable quantum computer [4].

A knowledge of the hyperfine transition frequencies of $^{31}$Si is necessary in order to assess the feasibility of laser cooling and trapping. This information may be deduced from the hyperfine structure and the isotope shift of the stable Si isotopes. Si has three stable isotopes: $^{28}$Si (92.23%), $^{29}$Si (4.67%) and $^{30}$Si (3.10%). The odd isotope, $^{29}$Si, has nuclear spin I=1/2. There are no previous hyperfine structure measurements on free $^{29}$Si atoms. One previous determination of the $^{28}$Si-$^{30}$Si isotope shift of the transition has been made by Holmes and Hoover using emissions from a hollow cathode lamp [5]. The present work provides a measurement of the hyperfine structure of $^{29}$Si and improves the accuracy of the $^{28}$Si-$^{30}$Si isotope shift by almost two orders of magnitude, in addition to providing an isotope shift for $^{29}$Si.

Hyperfine energy level splitting is a result of the interaction between valence electrons and the nucleus. The hyperfine A and B constants represent the interaction of the electromagnetic field produced at the nucleus by the electrons with the nuclear magnetic dipole moment, and with the nuclear electric quadrupole moment, respectively. The shift of a state with total angular momentum F from the unperturbed energy is given by [6]

$$\Delta E = \frac{A}{2}\left[F(F+1) - J(J+1) - I(I+1)\right] + \frac{B}{4}\frac{\left[\frac{3}{2}K(K+1) - 2J(J+1)I(I+1)\right]}{J(2J-1)I(2I-1)} \quad (1)$$

where J is the total electronic angular momentum, I is the nuclear spin, and K = F(F+1) – J(J+1) – I(I+1). The second term is absent for I=1/2 nuclei such as $^{29}$Si.

The relevant energy levels of Si are shown in Fig. 1. Figure 1(a) shows the fine structure of the ground state and the 221.7 nm transition. The nuclear spin-free $^{28}$Si and $^{30}$Si will have this single transition. The hyperfine structure of $^{29}$Si and the three hyperfine transitions studied in this work are shown in Fig. 1(b). The hyperfine structure of $^{31}$Si, based on the results of this work, is shown in Fig. 1(c). The laser cooling transition in each case is identified.

## II. EXPERIMENTAL METHOD

A schematic for the experimental apparatus is shown in Figure 2. Spectroscopy of the Si transition required a tunable deep ultraviolet (DUV) laser source at 221.7 nm. This was obtained by frequency doubling and doubling again of a Ti:Sapphire laser. A Nd:YVO$_4$ laser at 532 nm was used to pump a cw single frequency tunable Ti:Sapphire laser at 887 nm [7]. The IR laser light was frequency doubled to 443.5 nm in an external ring cavity utilizing an LBO crystal [8]. The output was doubled again in a second external ring cavity utilizing a BBO crystal. A DUV laser power of 2-25 mW was used in these experiments.

A beam of Si atoms was generated by a resistively heated effusive oven source with a 1 mm diameter aperture hole. The oven was machined from pyrolytic graphite with a screwed-on cap. The source and atomic beam were housed in a diffusion-pumped vacuum system at a base pressure of ~3x10$^{-7}$ Torr (~8x10$^{-7}$ Torr when the oven was operating). The oven was heated to approximately 1500°C, as determined by an optical pyrometer. A 1.4 mm x 1.4 mm aperture was placed 11 cm from the oven to collimate the Si atoms. In our experiment the lower energy state of the transition of interest is the metastable *3s$^2$3p$^2$ $^3$P$_2$*, which lies 223.157 cm$^{-1}$ above the *3s$^2$3p$^2$ $^3$P$_0$* ground state and is thermally populated. The DUV laser beam was introduced into the vacuum system through a fused silica Brewster window and passed perpendicularly to the Si beam at a distance of 15 cm from the collimating aperture. The laser beam waist diameter was approximately 2 mm. A photomultiplier tube (PMT) located above the interaction region was used to detect the fluorescence through an interference filter as the DUV laser frequency was scanned through the transition. The measured full width at half maximum of the Doppler broadened transition was ~70 MHz, in excellent agreement with the calculation based on the oven and collimator geometry [9]. For comparison, the natural linewidth of the transition is 8.8 MHz.

The frequency interval information was provided by sending a part of the 443.5 nm laser output into a temperature stabilized 1 m long confocal Fabry-Perot cavity with a free spectral range of 74.227(36) MHz [10]. The fluorescence signal from the photomultiplier as well as the fringe signal from the Fabry-Perot cavity were recorded simultaneously and digitized, as the laser frequency was scanned. The $^{28}$Si peak, two of the larger $^{29}$Si hyperfine peaks (F=5/2→7/2 and F=3/2→5/2), and the $^{30}$Si peak were visible in each scan, but the weak $^{29}$Si hyperfine component (F=5/2→5/2) required the addition of multiple spectra to be seen. Figure 3 shows the average of 24 scans, taken during a run on the same day. The smallest of the $^{29}$Si hyperfine peaks can be seen. Due to the drift in the Ti:Sapphire laser frequency, it was better to take each scan in a relatively short time (20 s per scan) rather than signal average for a long time.

## III. ANALYSIS AND RESULTS

The center of the $^{28}$Si peak was located for each scan, and 1-10 scans (adjacent in time) were averaged together after shifting the spectra to line up the $^{28}$Si peaks. This procedure improved the signal-to-noise ratio while compensating for the slow laser frequency drift. The Fabry-Perot cavity fringes were also averaged after applying the same shift corrections as for the Si spectra. The number of scans to be grouped together in a given run was chosen to optimize the signal-to-noise ratio for fits to the two larger $^{29}$Si peaks, the $^{30}$Si peak and the fringes. It was based on the signal strength of each spectrum (varying mainly in the Si atomic beam flux) and on the drift rate in the interferometer fringes relative to the $^{28}$Si peak. If the drift rate was large, fewer scans were averaged. At this stage, the weakest $^{29}$Si peak (F=5/2→5/2) was still too small to allow a measurement. For each grouped spectrum, a quadratic least squares fit to the four fringes nearest each Si peak was used to correct for laser scan non-linearity and to determine the frequency of each Si peak for that particular group. The frequency shifts for the three largest peaks relative to the $^{28}$Si peak and 1σ standard deviations for each data set were obtained by a weighted average and a weighted standard deviation of the mean of the group values. Once the larger peaks were analyzed for each group, the group spectra were adjusted to line up the $^{28}$Si peak, and an averaged spectrum for the whole data set was obtained, e.g., Fig. 3. It was then possible to fit the weakest peak and determine its frequency shift.

Five sets of data, each the result of 24-100 spectra, were obtained and are listed in Table 1. The hyperfine constant and the isotope shifts determined from the peak positions are given in Table 2. The previously published isotope shift result is given for comparison.

The present work represents the first measurement of the hyperfine A constant of $^{29}$Si. It provides the first measurement of the $^{29}$Si-$^{28}$Si isotope shift and a significant improvement in the accuracy of the $^{30}$Si-$^{28}$Si isotope shift. The uncertainties reported with our parameter values are 1σ deviations from the mean. The dominant contributions to the error limits are the noise of the laser-induced fluorescence signal and random fluctuations in the frequency scan of the laser. Systematic error due to frequency scan nonlinearity has been corrected and does not contribute significantly to the overall uncertainties.

IV. DISCUSSION

In light elements with small nuclear volumes, the field shift contribution to the isotope shift is generally small compared to the normal mass and specific mass shifts. The latter two effects are proportional to the difference in 1/reduced mass. To check this relationship for Si, the isotope shift for $^{29}$Si-$^{28}$Si is calculated from the measured $^{30}$Si-$^{28}$Si isotope shift. The calculated result, 1740 MHz, differs from the measured value, 1753.3(1.1) MHz, by only 13 MHz, confirming that field shifts are indeed small in this element. On this basis, the isotope shift for $^{31}$Si relative to $^{28}$Si can be predicted using reduced masses only. The result is given in Table 2 as 4876(13) MHz. An uncertainty equal to the difference of calculated and experimental isotope shifts for $^{29}$Si is assigned.

The hyperfine structure for $^{31}$Si may be deduced as well. For a given transition in a given element, the hyperfine A constants of different isotopes i and j are generally related to better than 1% by [11]

$$A_i = \left(\frac{I_j}{I_i}\right)\left(\frac{\mu_i}{\mu_j}\right)A_j \qquad (2)$$

Here, $\mu_i$, $\mu_j$ are the nuclear magnetic moments and $I_i$, $I_j$ are the nuclear spins of the two isotopes. The magnetic moment of $^{29}$Si (I=1/2) is well known, -0.55529$\mu_N$ [12]. The magnetic moment of $^{31}$Si (I=3/2) has not been measured directly, but a value of +0.55$\mu_N$ has been deduced by Webb et al. from a deformed nucleus model of the low-lying levels of $^{31}$Si [13]. This value is close to that measured for isotonic nuclei with the unpaired neutron in the same state, +0.64382$\mu_N$ and +0.633$\mu_N$ for $^{33}$S and $^{35}$Ar, respectively [12]. The agreement of the calculated quadrupole moment and deformation for $^{31}$Si in the Webb et al. paper with measurements for $^{33}$S provides further confirmation that the $^{31}$Si magnetic moment value is reasonable. A shell model calculation utilizing three different methods gave values of +0.35-1.03$\mu_N$ for $^{31}$Si [14]. Using the Webb et al. value and assuming an uncertainty of 0.1$\mu_N$, the A hyperfine constants for $^{31}$Si predicted using Equation (2) are given in Table 2. There will also be a B hyperfine constant for these states in $^{31}$Si, which cannot be determined by this method. A rough guess based on comparison to similar atoms indicates that the B hyperfine contribution is likely to be comparable to or less than the uncertainty due to the error limits on the A constants.

With the isotope shifts and hyperfine constants A, and the natural linewidth of this transition (8.8 MHz), the Doppler-free absorption spectrum for the Si isotopes can be calculated. The spectrum of a Si sample with natural abundance and enriched with 1% $^{31}$Si is shown in Fig. 4. Note that the cooling transition for $^{31}$Si (7/2 to 9/2 in Fig. 1(c) and Fig. 4) is separated from all other lines and is the most blue-shifted transition. The predicted location of this transition is 5510(120) MHz with respect to $^{28}$Si, with the error limit due mainly to the uncertainty in the $^{31}$Si magnetic moment. Laser cooling occurs with a slightly red-detuned laser from resonance. Thus only $^{31}$Si will be cooled and trapped, while all other isotopes would be heated. This graph also shows that the stable isotopes $^{28}$Si and $^{29}$Si may be cooled with little or no interference from other isotopes.

V. CONCLUSIONS

The *3s$^2$3p$^2$ $^3$P$_2$ → 3s3p$^3$ $^3$D$^o_3$* transition in silicon is investigated. The hyperfine A constants for $^{29}$Si are measured for the first time. The isotope shifts of the transition are measured and improved over the previous measurement. These measurements allow for a prediction of the A constants and isotope shifts for $^{31}$Si. The level structure is favorable for the laser cooling and trapping of $^{31}$Si.

ACKNOWLEDGEMENT: This work was supported by the W. M. Keck Foundation. We wish to acknowledge the help of William Czajkowski and Katherine Zaunbrecher in the experiment.

REFERENCES

[1]  B. E. Kane, "A silicon-based nuclear spin quantum computer"  Nature **393**, 133-137 (1998).
[2]  S. B. Hill and J. J. McClelland, "Atoms on demand:  Fast, deterministic production of single Cr atoms"  Appl. Phys. Lett. **82**, 3128 (2003).
[3]  J. L. Hanssen, J. J. McClelland, E. A. Dakin and M. Jacka, "Laser cooled atoms as a focused ion-beam source", Phys. Rev. A **74**, 063416 (2006).
[4]  W. M. Fairbank, Jr. and S. A. Lee, to be published.
[5]  J. R. Holmes and M. E. Hoover, "Isotope shift in first spectrum of silicon (Si I)", J. Opt. Soc. Am. **52**, 247 (1962).
[6]  G. T. Emery, "Hyperfine structure", in *Atomic, Molecular, and Optical Physics Handbook*, Ed. G.W.F. Drake, 198-205 (American Institute of Physics, Woodbury, NY, 1996)
[7]  The pump laser was a Coherent Nd:YVO$_4$ Verdi$^{TM}$ laser.  The tunable Ti:Sapphire laser was a Coherent MBR$^{TM}$ laser.
[8]  The doubling stages were Spectra Physics Wavetrains$^{TM}$.
[9]  N. F. Ramsey, "Molecular Beams", Sec. II.2.2 (Oxford University Press, London, 1956).
[10]  L. Hlousek, S. A. Lee, and W. M. Fairbank, "Precision wavelength measurements and new experimental lamb shifts in helium", Phys. Rev. Lett. **50**, 328-331 (1983)
[11]  S. Büttgenback, *Hyperfine Structure in 4d- and 5d-Shell Atom* (Springer-Verlag, Berlin, 1982) p. 30.
[12] *The 8th edition of the Table of Isotopes*, Ed. by Richard B. Firestone and Virginia S. Shirley (John Wiley & Sons 1996, 1998 and 1999) Appendix E.  The nuclear magnetic moments are also listed online at http://ie.lbl.gov/toipdf/mometbl.pdf.
[13]  V. H. Webb, N. R. Roberson and D. R. Tilley, "Analysis of the low-lying level structure of Si$^{31}$", Phys. Rev. **170**, 989 (1968).
[14]  P. W. M. Glaudemans, P. M. Endt and A. E. L. Dieperink, "Many-particle shell model calculation of electromagnetic transition rates and multipole moments in A=30-34 nuclei", Ann Phys. **63**, 134 (1971).

Table 1. Measured positions of the $^{29}$Si and $^{30}$Si lines in MHz relative to the $^{28}$Si line.

| Data Set | $^{29}$Si (5/2-7/2) MHz | $^{29}$Si (3/2-5/2) MHz | $^{29}$Si (5/2-5/2) MHz | $^{30}$Si MHz |
|---|---|---|---|---|
| 1 | 1114.6(1.3) | 2578.2(1.4) | 2980.0(7.7) | 3359.0(1.2) |
| 2 | 1119.1(6.3) | 2569.8(8.4) |  | 3357.9(6.6) |
| 3 | 1118.0(2.3) | 2576.3(2.4) | 2983.5(12.4) | 3359.5(1.8) |
| 4 | 1114.6(1.8) | 2576.8(5.7) | 2965.6(15.7) | 3359.5(1.8) |
| 5 | 1116.1(1.1) | 2573.7(2.5) | 2975.9(5.2) | 3361.0(1.2) |
| Weighted avg. | 1115.6(0.7) | 2576.8(1.5) | 2977.1(3.0) | 3359.9(0.6) |

Table 2. Isotope shifts and hyperfine structure constants observed for the stable isotopes and predicted for $^{31}$Si. Predictions from the measured values are given in italics.

| Parameter | $^{28}$Si | $^{29}$Si | $^{30}$Si | $^{31}$Si predicted |
|---|---|---|---|---|
| A ($3s^23p^2\ ^3P_2$) |  | -160.1(1.3) MHz |  | *53(10) MHz* |
| A ($3s3p^3\ ^3D°_3$) |  | -531.8(0.9) MHz |  | *176(32) MHz* |
| Isotope shift | 0 MHz | 1753.3(1.1) MHz | 3359.9(0.6) MHz |  |
| Isotope shift predicted from reduced mass |  | *1740 MHz* |  | *4876(13) MH* |
| Previous measurement |  |  | 3418(30) MHz[a] |  |

[a]Reference 5.

Fig. 1 (Color on line). Energy levels of silicon (not to scale). (a) Fine structure levels, (b) hyperfine structure of the transition for $^{29}$Si and (c) hyperfine structure energy levels of $^{31}$Si. The cooling transition is shown for each case.

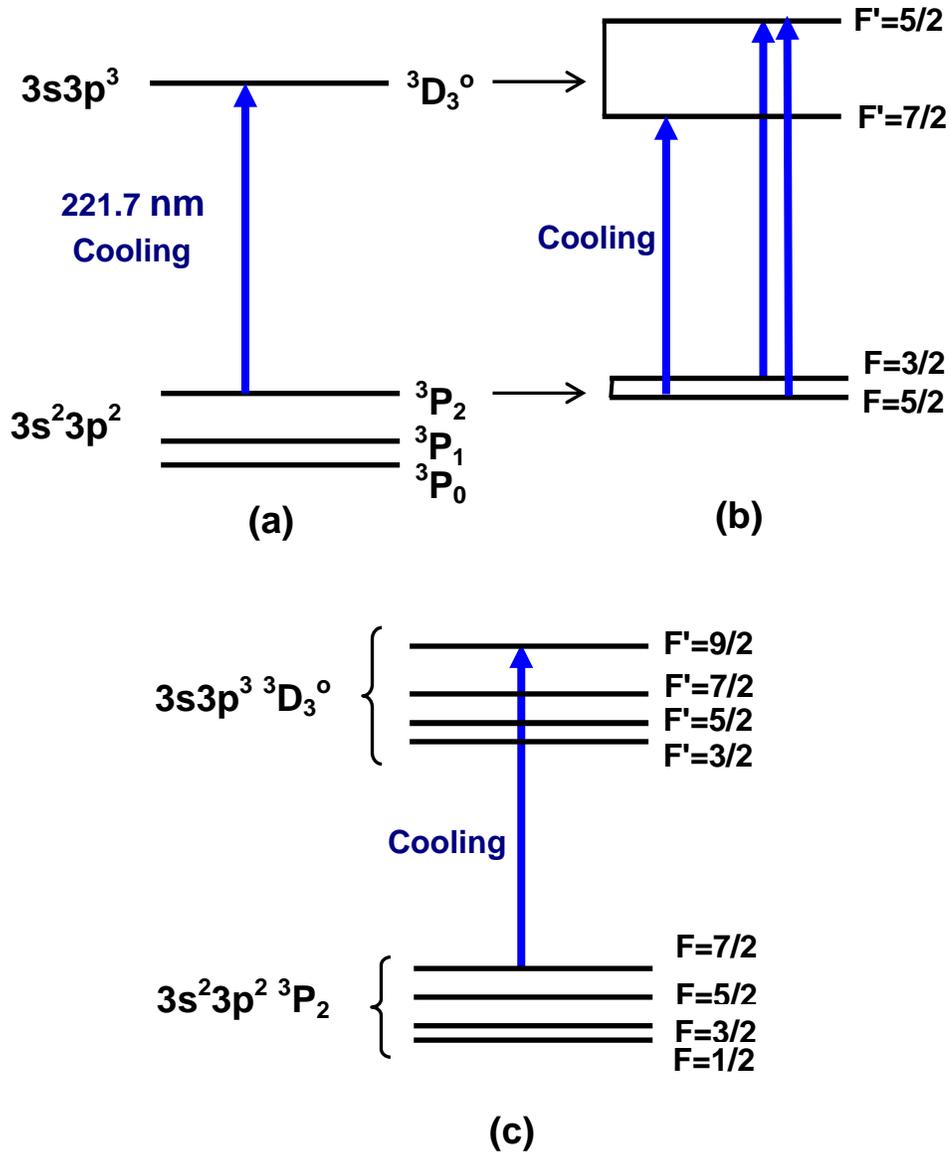

Fig. 2. (Color online). Schematic of the experimental apparatus for Si spectroscopy.

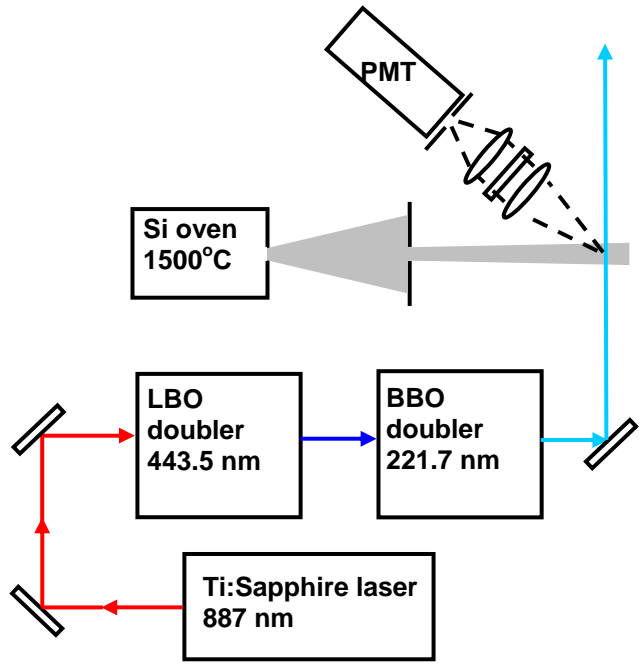

Fig. 3. (Color online). Averaged spectrum of Si for data set 1. The upper red trace is amplified by 20 times to show the smaller $^{29}$Si and $^{30}$Si peaks.

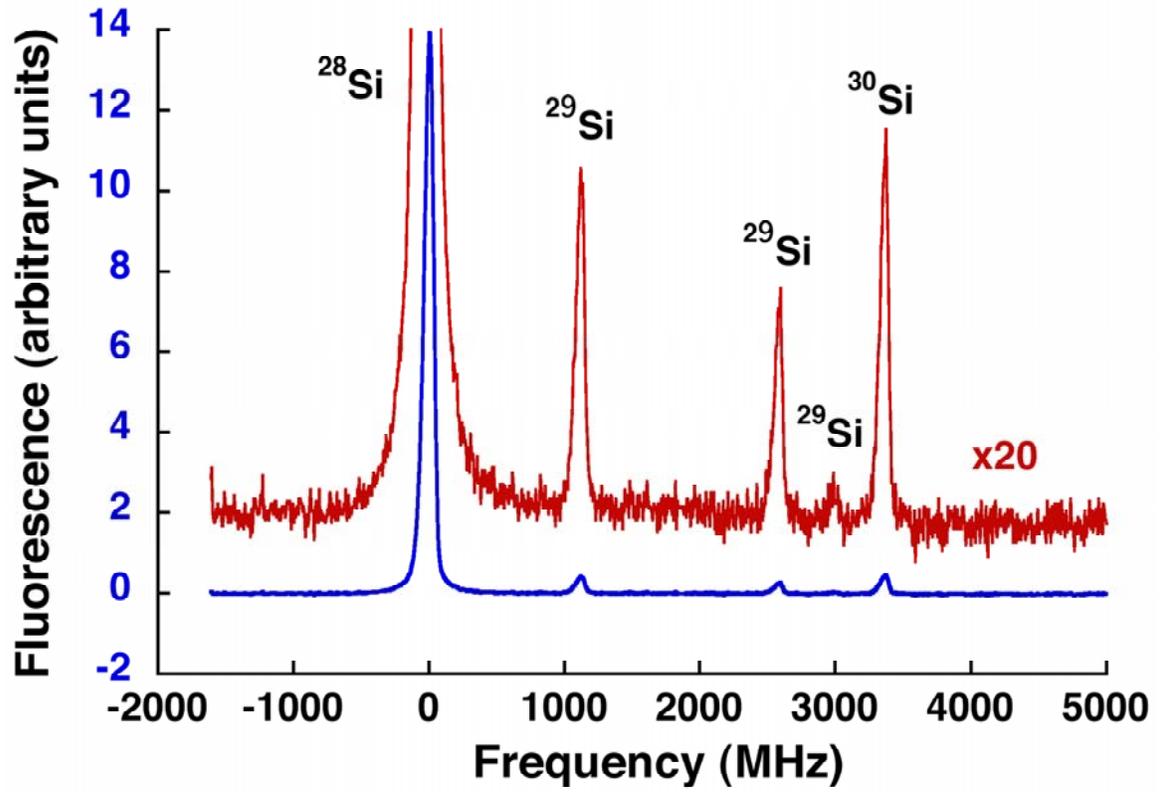

Fig. 4. (Color online). Calculated absorption spectrum of natural Si with 1% $^{31}$Si. The arrows show the approximate frequencies for laser cooling and trapping of the different isotopes.

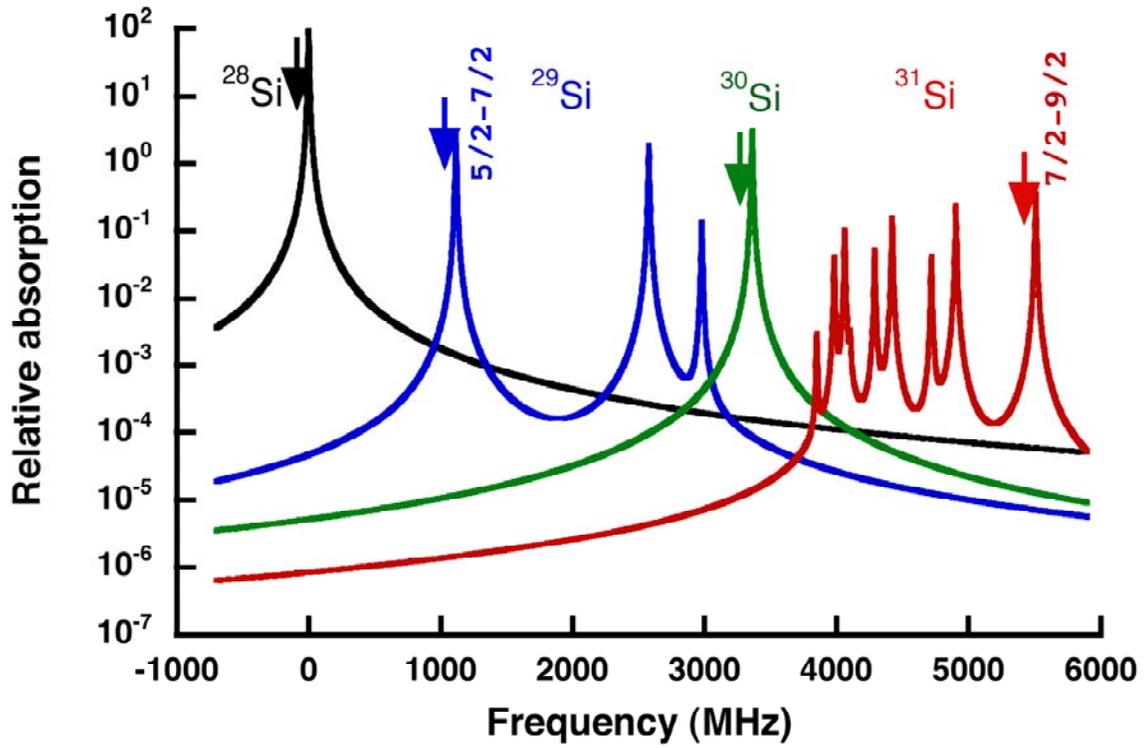